\documentclass[12pt]{article}
\usepackage{amsmath,amssymb}
\usepackage{epsfig}
\usepackage{graphicx}
\usepackage{a4}
\usepackage{latexsym}
\usepackage{cite}

\usepackage{color}
\usepackage{colordvi}

\DeclareGraphicsExtensions{.eps,.ps}

\usepackage{pslatex}
\usepackage[latin1]{inputenc}
\usepackage[T1]{fontenc}

\begin{document}

\begin{titlepage}
\noindent
DESY 10-164
\vspace{1.3cm}

\begin{center}
  {\bf
    \Large
    Parton Distribution Uncertainties using  Smoothness Prior \\
  }

  \vspace{1.5cm}
  {\large
    A.~Glazov$^{\,a}$, S.~Moch$^{\,b}$ and V.~Radescu$^{\,c}$ \\
  }

  \vspace{1.2cm}
  {\it
    $^a$Deutsches Elektronensynchrotron DESY \\
    Notkestra{\ss}e 85 D--22607 Hamburg, Germany \\
    \vspace{0.2cm}
    $^b$Deutsches Elektronensynchrotron DESY \\
    Platanenallee 6, D--15738 Zeuthen, Germany \\
    \vspace{0.2cm}
    $^c$ Physikalisches Institut, Universit\"at Heidelberg \\
    Philosophenweg 12, D--69120 Heidelberg, Germany \\
  }

  \vspace{1.4cm}
  {\large\bf Abstract} \\
  \vspace{0.5cm}

  \parbox{14.25cm}{
    A study of the parameterization uncertainty at low Bjorken $x \le 0.1$ 
    for the parton distribution functions of the proton is presented.
    The study is based on the HERA I combined data using a flexible  
    parameterization form based on Chebyshev polynomials with and without an additional 
    regularization constraint. 
    The accuracy of the data allows to determine the gluon density 
    in the kinematic range of $0.0005 \le x \le 0.05 $ with a small parameterization uncertainty. 
    An additional regularization prior leads to a significantly reduced uncertainty for $x \le 0.0005$.
  }
\end{center}

\end{titlepage}

\section{Introduction}
\label{sec:intro}
The accurate knowledge of the parton distribution functions (PDFs) plays an important role for predictions 
of hard scattering cross sections at $pp$ and $p\bar{p}$ colliders. 
The latter are computed in the perturbative approach including higher order radiative corrections, 
e.g. at next-to-leading order (NLO), which results in reduced theoretical uncertainties.
Particular cross sections, such as Drell-Yan production of $W, Z$ bosons at the LHC are even 
calculated to next-to-next-to-leading order (NNLO), see~\cite{Anastasiou:2003ds,Catani:2009sm},  
and exhibit a small theoretical uncertainty of $\sim 2\%$.
For these processes, the accuracy of the prediction is presently limited by the uncertainties of the PDFs.

The PDFs being non-perturbative by definition can be determined from fits to data 
from DIS $e$- and $\nu$-scattering, and from Drell-Yan experiments. 
These fits are performed using the well-known QCD evolution equations at NLO and NNLO~\cite{%
Gribov:1972ri,Altarelli:1977zs,Curci:1980uw,Furmanski:1980cm,Moch:2004pa,Vogt:2004mw}. 
The data are provided at discrete values of Bjorken $x$ and absolute four momentum transfer squared $Q^2$ 
with their statistical and systematic uncertainties.
With this given input, the uncertainty of the PDFs due to experimental errors are estimated 
using Hessian~\cite{Pascaud:1995qs,Pumplin:2002vw} and Monte Carlo~\cite{Jung:2009eq} methods.
Additional theoretical uncertainties arise, e.g. from unknown higher orders in the evolution or 
the treatment and scheme choice for heavy flavor contributions~\cite{Aivazis:1993pi,Buza:1996wv}.
These need to be considered separately (see e.g.~\cite{Adloff:2003uh,Aaron:2009kv}).

PDF fits require an ansatz for the parameterization by a certain function of $x$ at the starting scale $Q^2_0$ of the evolution. 
Fitting of experimental data at discrete points with an, in general, arbitrary function is an ill-posed problem which requires regularization. 
Typically Regge-theory inspired parameterizations are used with a small number of parameters which implicitly
contain smooth and regular behavior requirements for the PDFs. 
For these parameterizations, it is difficult to estimate the PDF uncertainty arising 
from the choice of a particular ansatz. 
Alternatively, flexible parameterizations based on a neural net approach were used recently~\cite{Ball:2010de}. 
The number of parameters in this approach is determined by the data using an over-fitting protection technique
which is an implicit regularization. 

In this note, a new study of the parameterization uncertainty at low Bjorken $x<0.1$ is performed.
An explicit regularization prior is introduced which disfavors resonant-like behavior of PDFs at low $x$  
and the impact of the prior on the parameterization uncertainty is evaluated 
with particular emphasis on the gluon density in the range $0.0005 \le x \le 0.05$. 
We choose a flexible ansatz for the PDFs at low $x$ using Chebyshev
polynomials.  
The analysis is based on the combined HERA I data~\cite{h1zeus:2009wt}.

\section{QCD Analysis Settings}
\label{sec:set}

The QCD analysis presented here is performed using as a sole input the combined H1 and ZEUS data on neutral and charged
 current $e^\pm p$ scattering double-differential cross sections collected during the HERA I run period of 1994-2000~\cite{h1zeus:2009wt}.
The kinematic range of the data extends from $0.045<Q^2<30000$ GeV$^2$ and $0.000006<x<0.65$, however in the 
QCD fit analysis only data with $Q^2\ge Q^2_{\rm min}= 3.5$ GeV$^2$ are considered in order to minimize the non-perturbative
 higher twist effects.

 The QCD fit is performed within the framework of the QCDNUM program
 implemented at NLO in QCD~\cite{Botje:2010ay}
and using a Zero-Mass-Variable-Flavor scheme. The fit minimizes a 
$\chi^2$ function as specified in~\cite{h1zeus:2009wt}.
 The PDFs are parameterized at the starting scale of $Q_0^2=1.9$~GeV$^2$.
We use
a flavor decomposition similar to~\cite{Chekanov:2005nn} as follows: $xd_{val}$, $xu_{val}$, $x\Delta=x\bar{u}-x\bar{d}$, 
 $xS=2(x\bar{u}+x\bar{d}+x\bar{s}+x\bar{c}+x\bar{b})$ where $c$ and $b$ quark densities are zero at the scales below
their corresponding thresholds. 
The PDFs are evolved in $Q^2$ using the NLO equations in the massless $\overline{MS}$-scheme 
and the charm and beauty quark PDFs are generated by evolution for scales above the respective thresholds.
The renormalization and factorization scales are set to $Q^2$.

Since this study is focused on the low $x<0.1$ region, the set-up for the QCD analysis is special
if somewhat simplified compared to modern high precision determinations
of PDFs, see e.g.~\cite{Alekhin:2009ni,Martin:2009iq,Nadolsky:2008zw,h1zeus:2009wt}. 
At low $x$ the PDFs are dominated by the gluon and sea-quark densities while
at high $x$ the valence-quark densities give larger contribution.
Thus, regarding the functional form for the PDFs, 
standard Regge-theory inspired parameterizations are used for the valence quarks:
\begin{eqnarray}
xu_v(x) &=& A_{u_v} x^{B_{u_v}} (1-x)^{C_{u_v}}\,, \\
xd_v(x) &=& A_{d_v} x^{B_{d_v}} (1-x)^{C_{d_v}}\,.
\end{eqnarray}
The low-$x$ behavior of the valence densities is assumed to be the same for $u$ and $d$ quarks by setting $B_{u_v} = B_{d_v}$. 
The normalizations $A_{u_v}$ and $A_{d_v}$ are determined by the fermion number
 sum rules. Therefore the valence sector is described by  three 
parameters.

For the gluon and sea densities a flexible Chebyshev polynomials based parameterization is used.
The polynomials
use $\log x$ as an argument to emphasize the low $x$ behavior. 
The parameterization is valid for $x>x_{\rm min} = 1.7\times 10^{-5}$ which 
covers the $x$ range of the HERA measurements for $Q^2\ge Q^2_{\rm min}$.
The PDFs are multiplied
by $(1-x)$ to ensure that they vanish as $x\to 1$. The resulting parameterization form is 
\begin{eqnarray}
x g(x) &=& A_g \left(1-x\right) \sum_{i=0}^{N_g-1} A_{g_i} T_i \left(-\frac{\textstyle 2\log x - \log x_{\rm min} } {\textstyle \log x_{\rm min} } \right)\,, \label{eq:glu} \\
x S(x) &=& \left(1-x\right) \sum_{i=0}^{N_S-1} A_{S_i} T_i \left(-\frac{\textstyle 2\log x - \log x_{\rm min} } {\textstyle \log x_{\rm min} } \right)\,, \label{eq:sea} 
\end{eqnarray}
where $T_i$ denote Chebyshev polynomials of the first type and the sum over $i$ runs up to $N_{g,S}=15$ 
for the gluon and sea-quark densities.
The Chebyshev polynomials are given by the well-known recurrence relation:
\begin{eqnarray}
  T_0(x) &=& 1, \\ 
  T_1(x) &=& x, \\
  T_{n+1}(x) &=& 2xT_n(x)-T_{n-1}(x). 
\end{eqnarray}
The normalization $A_g$ is determined by the momentum sum rule.
The advantage of the parameterization given by Eqs.~(\ref{eq:glu}),~(\ref{eq:sea}) is that momentum
sum rule can be evaluated analytically. Moreover, already for $N_{g,S} \ge 5$ the fit quality
is similar to that of a standard Regge-inspired parameterization with a similar number of parameters.

The PDF uncertainties are estimated using the Monte Carlo technique~\cite{Jung:2009eq}.
The method consists in preparing replicas of data sets by allowing the central values of the cross sections to 
fluctuate within their systematic and statistical uncertainties taking into account all point-to-point correlations.
The preparation of the data is repeated for $N>100$ times and for each of
these replicas a complete NLO QCD fit is performed to 
extract the PDF set. The PDF central values and uncertainties are estimated using the mean values and 
root-mean-squared (RMS) over the PDF sets obtained for each replica.

\section{Choice of the Smoothness Constraint}
\label{sec:th}
Fitting an arbitrary function to a discrete number of measurements is an ill-posed problem which requires
regularization. 
This regularization should have a physical motivation and be flexible enough to
cover the space of solutions compatible with QCD. 
At low $x$, the sea-quark PDF closely corresponds to a measurement of the structure
function $F_2$ in a DIS process. For DIS at low $x$, the invariant mass of the hadronic final state $W$
is calculated from $Q^2$ and $x$ as
\begin{equation}
W = \sqrt{ Q^2 \frac{\textstyle 1-x}{\textstyle x}}.
\end{equation}

Experimentally, it is well-known that for low values of $W$ and $Q^2$ 
the structure function $F_2$ displays resonances~\cite{CLAS}. These resonances, however, disappear for 
high $W>5$~GeV. 
The smooth behavior of $F_2$ for high $W$ can be explained phenomenologically by
high particle multiplicity of the  hadronic final state. 
A prior which disfavors resonant structures in $W$, for $W$ exceeding a certain
value $W_{\rm min}$, has therefore a strong phenomenological motivation.
This prior can be introduced as an additional penalty to the likelihood function
for the PDFs which are {\it longer} in $W$. Note that a prior using the length in $W$ as opposed to the length in $x$ enhances
sensitivity to the low $x$ region.  
For the $\chi^2$ function the prior corresponds
to an extra penalty term of a form
\begin{equation}
\chi^2_{\rm prior} = \alpha \left[\int_{W_{\rm min}}^{W_{\rm max}}\sqrt{ 1 + (xf'(W))^2} dW - \left(W_{\rm max}-W_{\rm min}\right) \right],
\end{equation}
where $\alpha$ is the relative weight of this PDF-length prior and 
the PDF $xf = xg, xS$, respectively. The prior $\chi^2_{\rm prior}$ has a minimum for the shortest PDF in $W$ which 
corresponds to a condition for the derivative, $xf'(W)=0$.
In this case, $\chi^2_{\rm prior}=0$ holds irrespective of the value of $\alpha$. 
The total $\chi^2_{\rm tot}$ is given by the sum of the $\chi^2$,
for the data versus theory comparison, and the penalty term 
\begin{equation}
\chi^2_{\rm tot} = \chi^2 + \chi^2_{\rm prior}\,.
\end{equation}

We choose $W_{\rm max}=320$~GeV which is the maximum value achievable at HERA. 
To stay far away from the resonance region,
$W_{\rm min}=10$~GeV is used which for $Q^2=1.9$~GeV$^2$ corresponds to 
$x\approx 0.02$. The prior is applied to both gluon and sea-quark densities 
at the starting scale $Q^2_0=1.9$~GeV$^2$ of the evolution.

\section{Results}
\label{sec:cheb}

\begin{figure}
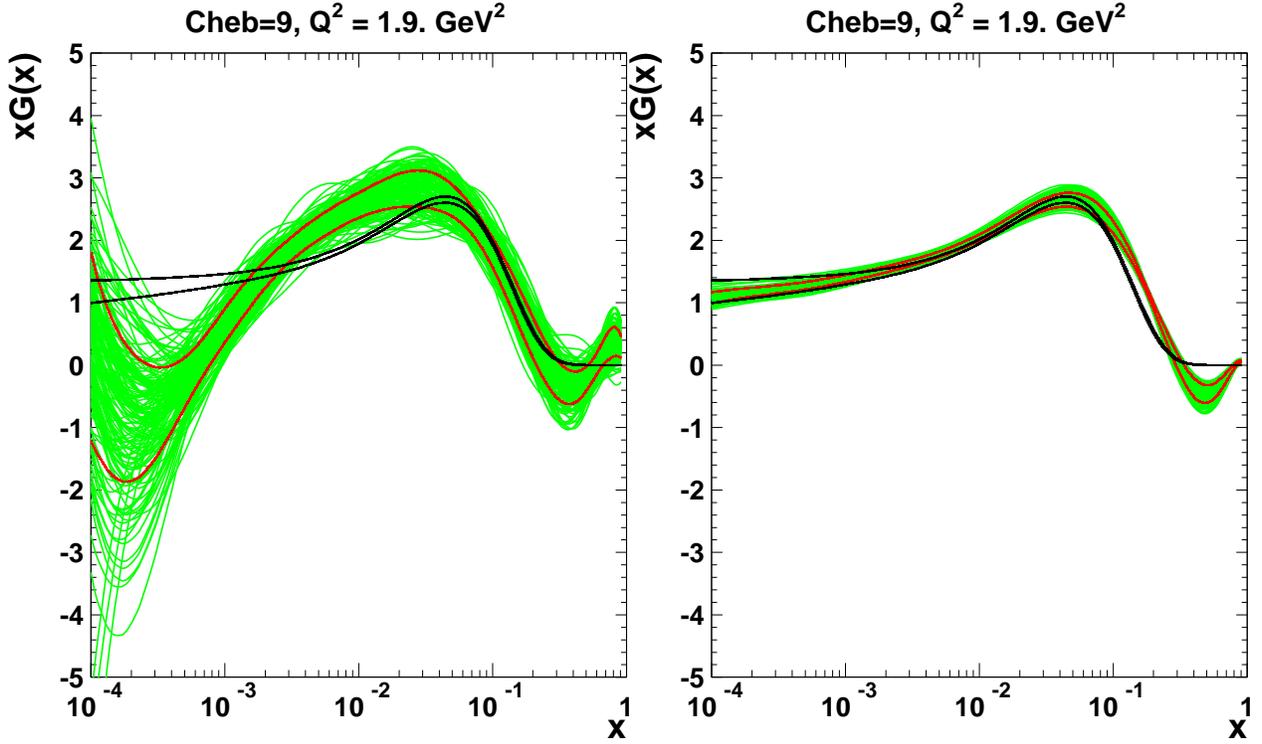

\centerline{\epsfig{file= glu9.epsi,width=0.5\linewidth}\epsfig{file= glu9.l5000.epsi,width=0.5\linewidth}}
\caption{\label{fig:glu15}
\small
The gluon PDF $xg(x)$ at the starting scale $Q^2_0=1.9$~GeV$^2$. The
green lines show fits to individual replicas of the data, 
the red lines show the RMS over the replicas. The black lines correspond to 
the error band of the gluon distribution using a standard parameterization and
it is to be compared to the case of the Chebyshev parameterization. On the left hand side, the gluon distribution
is shown using an unconstrained Chebyshev expansion to order nine, see Eq.~(\ref{eq:glu}), while on the right hand side the same distribution 
is displayed but with a tight length penalty $\alpha = 5000$~GeV$^{-1}$ applied.} 
\end{figure}

The Monte Carlo procedure of extracting PDFs is illustrated 
for $N_{par}\equiv N_g=N_S=9$ in Fig.~\ref{fig:glu15} which shows the gluon PDF 
at the starting scale $Q^2_0=1.9$~GeV$^2$ for each replica, with their RMS band.
The distributions are compared to those obtained using the standard parameterization form: $xg(x) = A_g x^{B_g}(1-x)^{C_g}$, 
$xS(x)= A_s x^{B_s}(1-x)^{C_s}$, where
$A_g$ is determined by the momentum sum rule.
The flexible parameterization does not suppress minima and maxima of the distribution as a function of $x$,
as a result, several of them are observed. 

\begin{figure}
\centerline{\epsfig{file= 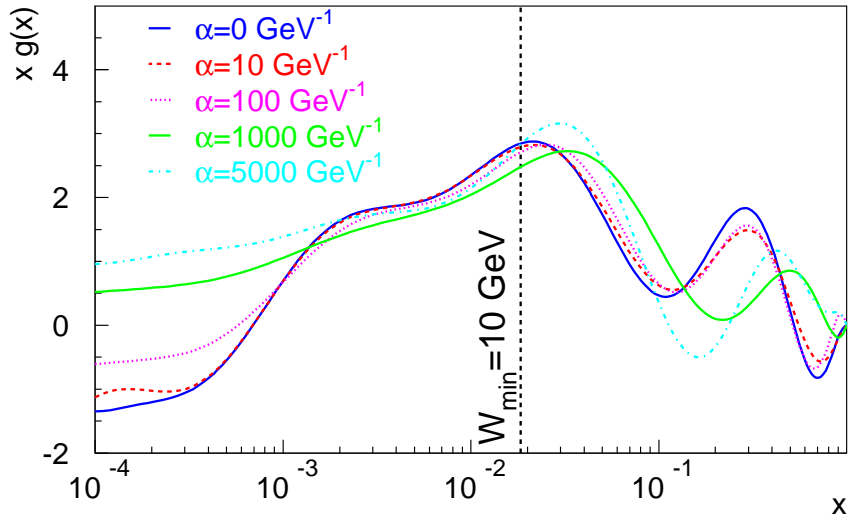,width=0.7\linewidth}}
\caption{\label{fig:leneffect} 
\small
The central value of the gluon PDF for various values of the length-prior 
weight $\alpha$ at the evolution starting scale $Q^2=1.9$~GeV$^2$ using the Chebyshev parameterization expanded to the 15$th$ order.
The vertical dashed line shows the $x$ value corresponding to the $W_{\rm min}=10$~GeV limit.}
\end{figure}
\begin{table}
\caption{\label{tab1}
\small
The quality of the fit in terms of $\chi^2/n_{\rm df}$ 
as a function of the length-prior weight $\alpha$.}
\begin{center}
\begin{tabular}{l|ccccc}
\hline
 $\alpha$, GeV$^{-1}$        &  0      & 10      &   100   & 1000  & 5000\\
 $\chi^2/n_{\rm df}$             & 560/557 & 561/557 & 572/557 & 626/557 & 767/557\\
\hline
\end{tabular}
\end{center}
\end{table}
Introducing the length prior to the fit  by changing the weight of the penalty term from $0$~GeV$^{-1}$ to $1000$~GeV$^{-1}$ 
increases the $\chi^2$ of the fit by $66$ units, see Table~\ref{tab1}. 
Further increase of the penalty term to $\alpha=5000$~GeV$^{-1}$ reduces
fit quality considerably with an additional increase of $\chi^2$ by $141$ units.  
For low values of $\alpha \le 100$, the impact of the penalty term on the
shape of the central value of the
gluon PDF is small while for $\alpha \ge 1000$ the distribution changes significantly, see
Fig.~\ref{fig:leneffect}. 
In addition,  the shape of the gluon distribution using a standard
parameterization can be reproduced by the Chebyshev parameterization of the gluon PDF if a tight length prior is applied to the fit, 
as demonstrated in Fig.\ref{fig:glu15}.

\begin{figure}
\centerline{
\epsfig{file= 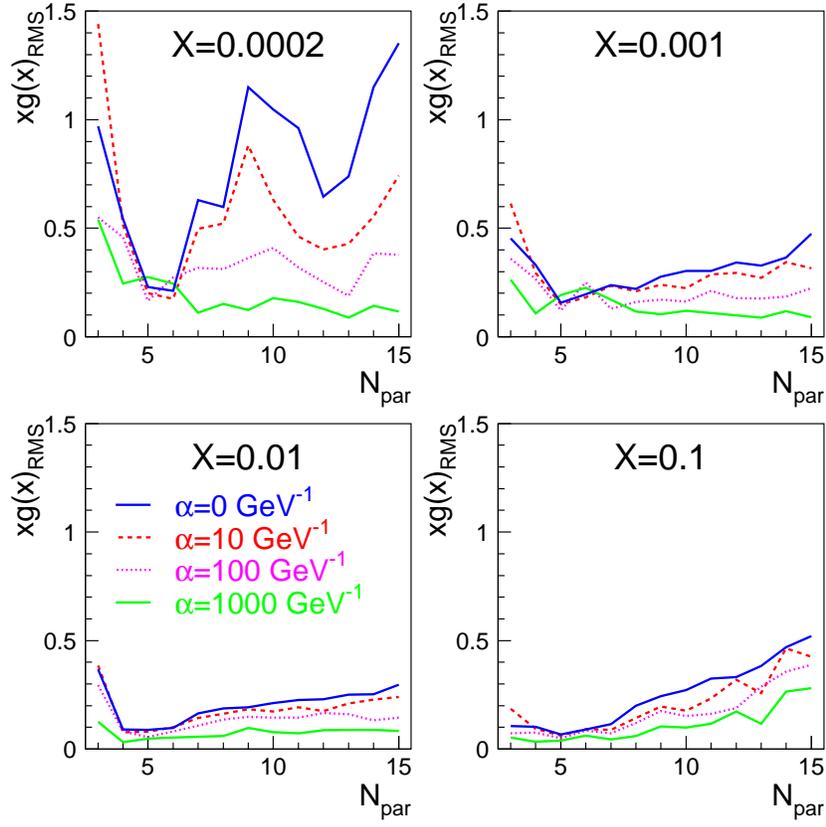,width=0.7\linewidth}}
\caption{\label{fig:gsl}
\small
The uncertainty of the gluon PDF as a function of 
${\rm N}_{\rm par} \equiv N_{g}=N_{S}$ for $Q^2=1.9$~GeV$^2$ at
fixed values of $x$ for different values 
of the length-prior weight $\alpha$ as
indicated by the figure's legend. 
}
\end{figure}

Fig.~\ref{fig:glu15} 
shows that the PDF uncertainty is very large for $x<0.0002$ and $x>0.05$ while for 
intermediate values of $x$ the data constrains the gluon PDF well. To quantify the dependence of the
uncertainty as a function of $N_{g,S}$, the RMS values at fixed values of $x$=$0.0002$, $0.001$, $0.01$
and $0.1$ are examined. 
The results of this investigation are summarized in Fig.~\ref{fig:gsl}.
For $x$=$0.01$ and $0.001$, the uncertainty stays approximately constant as the number
of parameters increases while for $x=0.0002$ and $x=0.1$ it increases significantly, if no penalty 
term is used.
The impact of the penalty term on the uncertainty of $xg(x)$ is large at low $x$.
Even low values of $\alpha\le 100$~GeV$^{-1}$ significantly reduce the uncertainty for $x=0.0002$
at large values of $N_{par}$, while for $x=0.1$ the penalty term has no impact. For the intermediate
$x$ values, the change of the uncertainty is moderate, indicating that for $p\le 100$~GeV$^{-1}$ 
the data provides stronger constraints on the gluon PDF than the prior.

\section{Summary}
\label{sec:sum}

The focus of this study has been on the parameterization uncertainty of PDFs at low $x$, 
especially of the gluon PDF in a fit to the HERA I data at NLO in QCD.
A flexible PDF parameterization based on Chebyshev polynomials has been chosen 
and the impact of an additional smoothness prior on the quality of the has been investigated.
We have found that the uncertainty of the fit is generally small in the $0.0005<x<0.05$ range. 
The uncertainty, however, increases significantly for larger and smaller $x$ values. 
The regularization with a smoothness prior, which disfavors resonant
structures for large values of $W$ allows to significantly reduce uncertainty also for the range $x<0.0005$.

\section*{Acknowledgments}
We are thankful to W.~Giele for suggesting the length prior and for 
valuable discussions about quantifying parton distribution function uncertainties~\cite{Giele:2001mr}.
We are also thankful to S.~Alekhin for interesting discussions. 
This work has been supported in part by Helmholtz Gemeinschaft 
under contract VH-HA-101 (Alliance {\it Physics at the Terascale}).

{\small

}

\end{document}